\documentclass[pra,twocolumn,tightenlines,showpacs,nofootinbib]{revtex4}
\usepackage{bm,dcolumn,amsmath,graphicx}

\newcommand{\eref}[1]{Eq.~(\ref{#1})}
\newcommand{\tref}[1]{Table~\ref{#1}}

\begin{document}
\title{Calculation of P,T-odd electric dipole moments for
       diamagnetic atoms $^{129}$Xe, $^{171}$Yb, $^{199}$Hg,
       $^{211}$Rn, and $^{225}$Ra}

\author{V. A. Dzuba$^1$}
\author{V. V. Flambaum$^1$}
\author{S. G. Porsev$^{1,2}$}
\affiliation{$^1$ School of Physics, University of New South Wales,
                  Sydney, NSW 2052, Australia}
\affiliation{$^2$ Petersburg Nuclear Physics Institute, Gatchina,
                  Leningrad district, 188300, Russia}

\date{\today}

\begin{abstract}

Electric dipole moments of diamagnetic atoms of experimental interest
are calculated using the relativistic Hartree-Fock and random-phase
approximation methods, the many-body perturbation
theory and configuration interaction technique. We consider P,T-odd
interactions which give rise to atomic electric dipole moment
in the second order of the perturbation theory. These include
nuclear Schiff moment, P,T-odd electron-nucleon interaction and
electron electric dipole moment. Interpretation of a new
experimental constraint of a permanent electric dipole moment
of $^{199}$Hg [W.~C.~Griffith {\it et al.}, Phys.~Rev.~Lett. {\bf 102},
101601 (2009)] is discussed.

\end{abstract}

\maketitle

\section{Introduction}
Search for a permanent electric dipole moment (EDM) of particles,
violating both parity and time-reversal invariance, has a long
history (see, e.g.,~\cite{KhrLam97}). The standard model predicts
tiny EDMs which cannot be detected at the present level of
experimental accuracy. However, different extensions of the standard
model (such as, e.g., supersymmetry) predict much
larger EDMs of the particles that, in principle, could be found using
the modern experimental technique. A reveal of such
EDMs would unambiguously lead to a contradiction with the
standard model.

According to \cite{HubPosRit07,PosRit05} previous limits on EDM produced
stringent constraints on electroweak bariogenesis and models of CP-violation,
but fall short of ruling out the simplest generic extensions of the standard model.
It was stated that the next generation of EDM experiments should be sufficiently
sensitive to provide a conclusive test.

A very significant step ahead in this direction has been done in a
recent work~\cite{GriSwaLof09}. The authors reported new result obtained
for a permanent EDM of $^{199}$Hg to be
$d(^{199}{\rm Hg}) = (0.49 \pm 1.29_{\rm stat} \pm 0.76_{\rm syst})
\times 10^{-29}\, e$ cm. Though the EDM is still consistent with zero,
its limit $|d(^{199}{\rm Hg})| < 3.1 \times 10^{-29}\, |e|$ cm is an
improvement of the previous Hg limit by a factor of 7.
Motivated by this result and implying future experimental progress we
have performed calculations of different
contributions to the atomic EDMs of $^{129}$Xe, $^{171}$Yb, $^{199}$Hg,
$^{211}$Rn, and $^{225}$Ra. For these atoms the experiments searching for
EDMs are underway.

The paper is organized as follows. In Sec.~\ref{Sec:GF} we discuss
different types of P,T-odd interactions that can lead to
an appearance of a permanent atomic EDM.
In Sec.~\ref{Sec:method} we describe the methods of calculations of
the EDMs. We start our calculations from the relativistic Hartree-Fock
method. Than we include many-body corrections using two different methods.
First, we apply a simple random-phase approximation (RPA) for the
closed-shell atoms. Second,
we apply the configuration interaction (CI) combined with the many-body
perturbation theory (MBPT)~\cite{DzuFlaKoz96b} approach to valence electrons
while use the RPA approach for the core.
Sec.~\ref{Sec:results} is devoted to an analysis and discussion
of the results. We present the results obtained for different contributions
to the atomic EDMs and compare them with other available data.
In Sec.~\ref{Sec:nucl} we discuss the neutron and proton
contributions to the total nuclear spin using the spherical shell model
of a nucleus. Sec.~\ref{Sec:concl} contains concluding remarks and
two final Tables where the recommended values of
the contributions to the EDMs of $^{129}$Xe, $^{171}$Yb, $^{199}$Hg,
$^{211}$Rn, and $^{225}$Ra are gathered and the limits on CP-violating
parameters based on new experimental limit for $^{199}\!{\rm Hg}$ are presented.
\section{General formalism}
\label{Sec:GF}
Our goal is to find the atomic EDM $d_{\rm at}$ defined as
${\bf d}_{\rm at} \equiv d_{\rm at} ({\bf F}/F)$,  where
${\bf F} = {\bf J}+{\bf I}$ with $\bf J$ being total angular momentum.
In this work we deal with the atoms with closed shells in their
ground states. In this case $\bf J = 0$, ${\bf F} = {\bf I}$,
and ${\bf d}_{\rm at} = d_{\rm at} ({\bf I}/I)$.

We consider several types of P,T-odd interactions
between particles leading to an appearance of an atomic EDM.
We restrict ourselves to contributions to the EDM
which occur in the second order of the perturbation theory.
The EDM induced in an
atomic state $|0\rangle$ due to an admixture of opposite-parity states that
appears in the second order of the perturbation theory can be written as
\begin{equation}
{\bf d}_{\rm at} = 2\, \sum_K \frac
{\langle 0|{\bf D}| K \rangle  \langle K|H| 0 \rangle}
{E_0 - E_K} ,
\label{Eq:d}
\end{equation}
where ${\bf D} = -|e|\,{\bf r}$ is the electric dipole operator,
$e$ is the electron charge, and $E_i$ are the energies of the states.

We will consider below:
1) the tensor-pseudotensor P,T-odd electron-nucleon ($e$-$N$) interaction,
2) the pseudoscalar-scalar P,T-odd $e$-$N$ interaction,
3) the nuclear Schiff moment, and
4) the interaction of the electron EDM with the internal nuclear
magnetic field of the atom.

 It is worth noting that the operators describing
all these interactions have certain similar features. All of them 1) are
proportional to the nuclear spin {\bf I} and 2) have strong singularity.
The similar nature of these operators leads to a suggestion that the calculations
depend very little on their specific form. It allows us
to expect that the relative contributions of different many-body parts of
calculations will remain approximately the same for all operators.
As we will show below this suggestion is fully justified.

\subsection{Electron-nucleon P,T-odd interactions}
\label{Subsec:e_N}
We start our consideration with brief reminder
of the main features of the $e$-$N$ P,T-odd interaction leading
to appearance of atomic electric dipole moments in the second order
of the perturbation theory.
A detailed description can be found elsewhere~\cite{KhrLam97,GinFla04}.
This interaction has the following form (see, e.g.,~\cite{GinFla04}):
\begin{eqnarray}
H &=&  \frac{G}{\sqrt{2}} \sum_N \,
\left[
 C^N_T\, \overline{N} i \gamma_5 \sigma_{\mu \nu} N \,
          \overline{e}   \gamma_5 \sigma_{\mu \nu} e \right.\nonumber \\
 &+&  \left. C^N_P\, \overline{N} i \gamma_5 N \, \overline{e} e \right] .
\label{Eq:H}
\end{eqnarray}
Here $C^N_T$ and $C^N_P$ are dimensionless coupling constant
characterizing tensor-pseudotensor and pseudoscalar-scalar
P,T-odd electron-nucleon interactions for the nucleon $N$;
$\sigma_{\mu\nu} = (\gamma_\mu \gamma_\nu - \gamma_\nu \gamma_\mu)/2$,
$\gamma_5$ and $\bm \gamma$ are the Dirac matrices:
$\gamma_5 = \left(
\begin{array}{rr}
 0 & -1 \\
-1 &  0
\end{array} \right) $ and
${\bm \gamma} = \left(
\begin{array}{rr}
      0       & {\bm \sigma} \\
-{\bm \sigma} &      0
\end{array} \right) $.

In accordance with \eref{Eq:H} we can represent the Hamiltonian $H$ as
$H \equiv  H_T +H_P$, where in the coordinate representation
(atomic units $\hbar = m_e = |e| = 1$ are used throughout)
\begin{eqnarray}
H_T = i \sqrt{2}\, G\,
C_T\, {\bm \gamma} \langle {\bm \sigma}_N \rangle\, \rho(r) ,
\label{Eq:H2}
\end{eqnarray}
\begin{eqnarray}
H_P = -\frac{G}{\sqrt{2}}\, \frac{1}{2 m_p c}\, C_P\,
\gamma_0\, {\bm \nabla} \rho(r)\, \langle {\bm \sigma}_N \rangle .
\label{Eq:H3}
\end{eqnarray}
Here $G$ is the Fermi constant,
$c$ is the speed of light
(in atomic units $c = 1/\alpha \approx 137$) and $m_p$ is the nucleon mass.

We denote
$$C_T \langle {\bm \sigma}_N \rangle \equiv
\left\langle C^p_T \sum_p {\bm \sigma}_p + C^n_T \sum_n {\bm \sigma}_n \right\rangle,$$
$$C_P \langle {\bm \sigma}_N \rangle \equiv
\left\langle C^p_P \sum_p {\bm \sigma}_p + C^n_P \sum_n {\bm \sigma}_n \right\rangle,$$
where $\langle ... \rangle$ means averaging over the nuclear state
with the nuclear spin ${\bm I}$.

In \eref{Eq:H3} we keep only the term in the lowest nonvanishing
approximation in $m_p^{-1}$. Note that in this equation the operator
$\bm \nabla$ acts only to $\rho(r)$, where $\rho(r)$ is the nuclear
density distribution.

Since we are dealing with very singular operators the model of
the nuclear density distribution can be important. To check
this point we have carried out calculations for two models. In one of them
the nucleus was treated as a charged sphere with the radius $R$, i.e.,
\begin{equation}
\rho(r) = \frac{3}{4\pi R^3}\, \theta(R-r) .
\label{Eq:rho}
\end{equation}
In other model it was used the Fermi distribution
\begin{equation}
\rho(r) = \frac{\rho_0}{1+\exp{\frac{r-R}{a}}} .
\label{Eq:rho1}
\end{equation}
where $\rho_0$ is the normalization parameter determined by
$\int \rho\, dV = Z$.
We have found that the results obtained for each of these
models were numerically very close to each other.

Similar to the expression for $H$ it is convenient to represent ${\bf d}_{\rm at}$
as a sum of two terms ${\bf d}_{\rm at} = {\bf d}_{\rm at}^T + {\bf d}_{\rm at}^P$,
where ${\bf d}_{\rm at}^T$ and ${\bf d}_{\rm at}^P$ correspond to the operators
$H_T$ and $H_P$ given by Eqs.~(\ref{Eq:H2}) and (\ref{Eq:H3}).
Explicit expressions for ${\bf d}^T_{\rm at}$ and ${\bf d}^P_{\rm at}$ can be derived
from \eref{Eq:d} by replacing the operator $H$ to $H_T$ and $H_P$,
correspondingly.

It is convenient to determine the quantities $d^T_{\rm at}$ and $d^P_{\rm at}$
as follows:  ${\bf d}^{T,P}_{\rm at} = d^{T,P}_{\rm at} \langle {\bm \sigma}_N \rangle
\sim d^{T,P}_{\rm at}\, {\bf I}/I $.
The coefficient of proportionality in this expression depends on
a model of the nucleus. An accurate treatment of the nuclear structure is
beyond the topic of this work. For a spherical
shell model of the nucleus this coefficient can be easily
found for different atoms. We will discuss this problem in more detail
in Sec.~\ref{Sec:nucl}.
\subsection{The nuclear Schiff moment}
The Schiff moment is a nuclear moment violating both parity and
time-reversal invariance. It is caused by $P,T$-odd
nuclear forces and it takes into account screening of external electric
field by atomic electrons. We use the form of the Hamiltonian for the
interaction of atomic
electrons with the nuclear Schiff moment suggested in Ref.~\cite{FlaGin02}:
\begin{equation}
 H_{\rm SM} = -\frac{3{\bf S}{\bf r}}{B}\,\rho(r),
\label{Eq:SM}
\end{equation}
where $B \equiv \int\rho(r)\,r^4 dr$ and ${\bf S}$ is the
Schiff moment vector defined as ${\bf S} = S\,({\bf I}/I)$
with $S$ being the coupling constant.

Its contribution to the EDM of the closed-shell atoms
has been considered in detail in~\cite{DzuFlaGin02_EDM,DzuFlaGin07}.
As it follows from ~\cite{DzuFlaGin02_EDM} the results obtained
in the frame of multiparticle approach combining
the CI with the MBPT agreed (within 10\%)
with the results obtained by the RPA method.
Here we perform similar calculations mostly for consistency test
and for completeness.
\subsection{The electron EDM}
\label{Subsec:de}
An interaction of the electron EDM $d_e$ with the electromagnetic
field strength $F_{\mu\nu}$ can be written in the relativistically
covariant form as
\begin{equation}
H_e = \frac{d_e}{2} \overline{\psi} \gamma_5 \sigma_{\mu\nu} \psi F_{\mu\nu} .
\label{Eq:Hd}
\end{equation}
Here $\overline{\psi} = \psi^\dagger \gamma_0$ and $\psi$ is
determined in Appendix~\ref{Ap:1}.

Again we will consider here only effect appearing
in the second order of the perturbation theory.
It characterized by an interaction of the electron EDM
with the magnetic field ${\bf B}$ created by the nuclear magnetic moment.
The operator of this interaction ($H_B$) can be written as
\begin{equation}
 H_B = -i d_e {\bm \gamma} {\bf B} .
\label{Eq:He10}
\end{equation}
The magnetic field ${\bf B}$ can be represented by
\begin{eqnarray}
{\bf B} &=& \nabla \times \frac{({\bf M} \times {\bf r})}{r^3_>} \\
&=& \frac{3 ({\bf M}{\bf n}){\bf n} - {\bf M}}{r^3}\, \theta(r-R)
+ \frac{2 {\bf M}}{R^3}\, \theta(R-r) , \nonumber
\label{Eq:B0}
\end{eqnarray}
where
$ \theta(x) =
\left\lbrace
\begin{array}{l}
1,\, x \geq 0  \\
0,\, x < 0
\end{array}
\right. $
and ${\bf n} \equiv {\bf r}/r$.
The contribution to the atomic EDM,
can be written as ${\bf d}_{\rm at}^B = d_{\rm at}^B ({\bf I}/{I})$ and
found from \eref{Eq:d} by replacing $H \rightarrow H_B$.

\section{Methods of calculation}
\label{Sec:method}
\subsection{RPA for the closed shells}
\label{SubSec:one-el}
Here we describe  a simple method suitable for calculations of atomic EDM
for atoms with closed shell. On the first stage we solve
Dirac-Hartree-Fock (DHF) equations in the $V^N$ approximation (i.e., including
all electrons forming the ground state of the atom in a self-consistency
procedure).
\begin{equation}
 \hat H_0\, \psi_c = \varepsilon_c \,\psi_c .
\end{equation}
Here $H_0$ is the relativistic Hartree-Fock Hamiltonian and
$\psi_c$ and $\varepsilon_c$ are single-electron wave functions
and energies.

At the next step we construct virtual orbitals.
Different techniques can be used for this procedure. One approach is to multiply the
previous orbital of the same partial wave to a smooth function of $r$
with subsequent orthogonalization of this orbital to all the rest orbitals.
This method was described in detail in Refs.~\cite{KozPorFla96,Bog91}.

Another method is to construct a basis set using the
B-spline technique developed at the University of Notre Dame
\cite{JohBluSap88}. We use 60 B-splines of order 9 in a cavity of radius
$R_{\rm max} = 30\, a_B$, where $a_b$ is the Bohr radius. This relatively large
number of B-splines is needed due to singularity of the P,T-odd operators.
This requires very detailed
description of the wave functions in the vicinity of the nucleus.

Further, we consider an atom in external field and solve the
RPA equations (self-consistent DHF in an external field)
\begin{equation}
(\hat H_0 -\varepsilon_c)\delta \psi_c = -(\hat F + \delta V^{N}_F) \psi_c ,
\label{Eq:RPA}
\end{equation}
where $\hat F$ is the
operator of external field, $\delta V^{N}_F$ is the correction to the
self-consistent potential due to the effect of external field. Index
$c$ numerates single-electron functions ($\psi_c$) of the
closed-shell core. The RPA equations \eref{Eq:RPA} are solved
self-consistently for all states in the core for all external fields
involved in the problem.
\subsection{CI+MBPT}
\label{SubSec:many-el}
A more sophisticated and accurate way to calculate atomic EDMs is to use
the configuration interaction technique for valence electrons while still
use the RPA approach for the core. This would allow us to check the accuracy
of the RPA calculations. It is especially important for the
atoms having two external $s$ electrons: Yb, Hg, and Ra. The effect
of external electrons on different properties of these atoms
is large and accurate treatment of interaction between
them is needed.

We consider Yb, Hg, and Ra as atoms with two valence electrons above
closed-shell cores [$1s,...,4f^{14}$], [$1s,...,5d^{10}$], and
[$1s,...,6p^6$], respectively.
In this paper we follow approach
suggested in~\cite{DzuFlaKoz96b} which combines
the many-body perturbation theory with the configuration
interaction method. We will refer to it as the CI+MBPT formalism.
The MBPT is used to include excitations from the core into the effective
Hamiltonian for valence electrons. After
that the multiparticle relativistic 
equation for valence electrons
is solved within the CI framework to find the wave functions
and the low-lying energy levels.

In the CI+MBPT method, the energies and wave functions are
determined from the time-independent 
equation
\begin{equation}
H_{\rm eff}(E_n) \Phi_n = E_n \Phi_n,
\label{Phi}
\end{equation}
where the effective Hamiltonian is defined as
\begin{equation}
H_{\rm eff}(E) = H_{\rm FC} + \Sigma(E).
\label{Heff}
\end{equation}
Here $H_{\rm FC}$ is the Hamiltonian in the frozen core
approximation and $\Sigma$ is the energy-dependent correction, which
takes into account virtual core excitations. The operator $\Sigma$
completely accounts for the second-order perturbation theory over
residual Coulomb interaction.

Since we are interested in calculating the atomic EDMs, we need to construct
the corresponding effective operators for valence electrons
\cite{DzuKozPor98,PorRakKoz99P,PorRakKoz99J}. To do that we can extend
the concept of the effective Hamiltonian $H_{\rm eff}$ to other operators
such as the effective dressed electric-dipole operator D$_{\rm eff}$
and the P,T-odd operators. These operators
account for the core-valence correlations.
As in pure RPA approach of Sec.~\ref{SubSec:one-el}, we
solve the RPA equations summing a certain sequence of many-body diagrams
to all orders of MBPT~\cite{DzuKozPor98,KolJohSho82,JohKolHua83}.
Since requirements to the accuracy of calculations are not very
high we disregard in this consideration small corrections
like normalization and structural radiation.

We perform the calculations in the CI+MBPT method in $V^{N-1}$ and
$V^{N-2}$ potentials. The former is a bit more ``natural'' for the
lowest-lying odd-parity states of the considered atoms such
as $nsnp\, ^{3,1}\!P^o_1$ ($n=6$ for Yb and Hg, and $n=7$ for Ra),
because $6p_{1/2,3/2}$ (or $7p_{1/2,3/2}$) orbitals are constructed at the stage
of solving DHF equations for the configuration [core]$nsnp$. The latter
is somewhat simpler (e.g, due to an absence of the subtraction diagrams).
We have checked that the final results in both potetials are in good
agreement with each other.
For this reason when we discuss results, we do not distinguish between
these potentials.

Two different basis sets described in section \ref{SubSec:one-el} were
used in the $V^{N-1}$ and $V^{N-2}$ approximations. First
set~\cite{KozPorFla96,Bog91} was used in the $V^{N-1}$ and B-spline
set~\cite{JohBluSap88} was used in the $V^{N-2}$ calculations.
At the CI stage of the $V^{N-1}$ calculations the one-electron basis set for Yb included
1$s$--13$s$, 2$p$--13$p$, 3$d$--12$d$, 4$f$--11$f$, and 5$g$--7$g$
orbitals, where the core- and 5$d$ and 6$p$
orbitals are Dirac-Hartree-Fock ones and all the rest are the
virtual orbitals. For Hg and Ra the basis sets were insignificantly larger.
In all cases the basis sets were numerically complete and the full CI
was made for two valence electrons.
At the stage of the MBPT calculations a more extended basis sets,
including more basis functions, were used.
For instance, for Yb it included 1$s$--26$s$, 2$p$--26$p$, 3$d$--25$d$,
4$f$--17$f$, and 5$g$--12$g$ orbitals.

All B-splines up to $l_{max}=5$ were used to calculate $\Sigma$
in the $V^{N-2}$ approximation. 18 lowest basis states above the core
in each parial wave up to $l_{max}=3$ were used on the CI stage of
the $V^{N-2}$ calculations.
\section{Results and discussion}
\label{Sec:results}
\subsection{RPA for the closed shells}
\label{SubSec:one-el_res}
In the frame of the RPA method  discussed
in Sec.~\ref{Sec:method} we have performed the calculations of different
contributions to the EDMs of the diamagnetic atoms presented in
~\tref{T:atom}. For these atoms the experiments searching for
the EDMs are planned or are underway.
\begin{table}
\caption{The isotopes of the atoms considered in this work.
$\mu$ are the magnetic moments expressed in nuclear magnetons.}

\label{T:atom}

\begin{ruledtabular}
\begin{tabular}{lcccc}
     &  Z &  A  &  I  & $\mu$ \\ 
\hline
 Xe  & 54 & 129 & 1/2 & -0.7778 \\
 Yb  & 70 & 171 & 1/2 &  0.4119 \\
 Hg  & 80 & 199 & 1/2 &  0.5059 \\
 Rn  & 86 & 211 & 1/2 &  0.60   \\
 Ra  & 88 & 225 & 1/2 & -0.734  \\
\end{tabular}
\end{ruledtabular}
\end{table}

We can rewrite \eref{Eq:d} as follows
\begin{equation}
{\bf d}_{\rm at} = 2 \sum_{a,k} \frac
{\langle k|{\bf r}| a \rangle  \langle k|H| a \rangle}
{\varepsilon_a - \varepsilon_k} ,
\label{Eq:one_el}
\end{equation}
where the summation is over the quantum numbers of the one-electron
core states ``a'' and excited states ``k'', and $\varepsilon_i$ are
the one-electron energies.

\subsubsection{P,T-odd $e$-$N$ interaction}
We start the discussion from tensor-pseudotensor and pseudoscalar-scalar
$e$-$N$ P,T-odd interactions.
Using the Wigner-Eckart theorem and going over to the reduced matrix elements
(ME) we obtain for the contributions $d_{\rm at}^T$ and $d_{\rm at}^P$ to atomic EDM
\begin{equation}
d_{\rm at}^T = \frac{2 \sqrt{2}\,G}{3}\, C_T \, \sum_{a,k} \frac
{\langle k||r|| a \rangle  \langle k|| i \gamma\, \rho(r)|| a \rangle}
{\varepsilon_k - \varepsilon_a} ,
\label{Eq:d2_1e}
\end{equation}
\begin{equation}
d_{\rm at}^P = -\frac{G\,C_P}{3 \sqrt{2}\, m_p c}\,
\sum_{a,k}
\frac {\langle k||r|| a \rangle \langle k|| n\, (d\rho/dr) \gamma_0|| a \rangle}
{\varepsilon_k - \varepsilon_c} .
\label{Eq:d3_1e}
\end{equation}
The explicit expressions for the reduced MEs of the P,T-odd operators in
Eqs.~(\ref{Eq:d2_1e}) and (\ref{Eq:d3_1e}) are given in Appendix~\ref{Ap:1}.

In~\tref{T:d2_1e} we present the values of $d_{\rm at}^T$ obtained for all
isotopes listed in \tref{T:atom} in pure DHF approximation and including RPA corrections.
Note that the RPA corrections must be included only for one operator
in~\eref{Eq:d2_1e} or~\eref{Eq:d3_1e}~\cite{Mar85}. In other words,
when we include the RPA corrections
for the electric dipole operator $r$ we must not include them for the P,T-odd operator
and {\it vice versa}. Certainly both approaches should lead to the same result.
It allows us to test consistency of the calculations.

Our results for Xe are in good agreement with other data. Our RPA
value for $^{199}$Hg is in excellent agreement with similar
calculations by M{\aa}rtensson-Pendrill~\cite{Mar85}
while the result obtained
in Ref.~\cite{LatAngCha08} is somewhat larger. There is also a reasonable
agreement between the result found in this work and the estimate
obtained by Sushkov, Flambaum and
Khriplovich from the analytically derived formula~\cite{SusFlaKhr84}. To the best
of our knowledge there are no other data for Yb, Ra, and Rn to compare with.

As is seen from \tref{T:d2_1e}, inclusion of the RPA corrections leads to increasing
atomic EDM. For the noble gases (Xe and Rn) the RPA corrections contribute at the level of
30\%, while for the atomic Hg, Yb, and Ra, which have two $s$ electrons above closed
shells, the RPA corrections are much larger. In fact, they increase the EDM
 2.5 times for Hg and 5 times for Yb and Ra as compared to the DHF
 results. The reason for this increase is that the two $s$ electrons
are loosely bound and can be easily excited. As a result, an account for
the higher orders of the perturbation theory (like the RPA corrections) leads
to significant change of the ``bare'' results obtained in the DHF approximation.

\begin{table}
\caption{The values of ${\bf d}_{\rm at}^T$
in units ($10^{-20} C_T\, \langle {\bm \sigma_N} \rangle |e|$ cm)
obtained in DHF and RPA approximations are presented.
The results are compared with other data.}
\label{T:d2_1e}
\begin{ruledtabular}
\begin{tabular}{lccccc}
                             & $^{129}$Xe & $^{171}$Yb & $^{199}$Hg & $^{211}$Rn & $^{225}$Ra \\
\hline
This work (DHF)              &    0.45    &   -0.70    &   -2.4     &    4.6     &   -3.5      \\
Ref.\cite{Mar85} (DHF)       &    0.41    &            &   -2.0     &            &             \\
Ref.\cite{DzuFlaSil85} (DHF) &    0.41    &            &            &            &             \\
Ref.\cite{SusFlaKhr84}       &    0.6     &            &   -3.9     &            &             \\[1mm]
This work (RPA)              &    0.57    &   -3.4     &   -5.9     &    5.6     &    -17      \\
Ref.\cite{Mar85} (RPA)       &    0.52    &            &   -6.0     &            &             \\
Ref.\cite{LatAngCha08} (RPA) &            &            &   -6.75    &            &             \\
\end{tabular}
\end{ruledtabular}
\end{table}
The results obtained for ${\bf d}_{\rm at}^P$ are listed in~\tref{T:d3_1e}.
In Ref.~\cite{FlaKhr85} Flambaum and Khriplovich suggested a method to establish
the correspondence between contributions of the tensor-pseudotensor
and pseudoscalar-scalar P,T-odd operators using the expressions
for the reduced MEs of these operators (see Appendix~\ref{Ap:1}).

This correspondence can be obtained using the properties of the wave
functions $f_s$, $g_s$, $f_{p_{1/2}}$, and $g_{p_{1/2}}$ in the vicinity
of the nucleus (see, e.g.,~\cite{FlaGin02}). For instance,
\begin{equation}
f_{p_{1/2}}(r) \approx g_{p_{1/2}}(R)\, \frac{1}{2}\,Z \alpha x
\left( 1-\frac{1}{5} x^2 \right),
\end{equation}
where $x \equiv r/R$.

At $r=R$ we obtain
\begin{equation}
f_{p_{1/2}}(r) \approx g_{p_{1/2}}(R)\, \frac{1}{2}\,Z \alpha
\times 0.8\,
\end{equation}
and after some transformations we find the correspondence
\begin{equation}
 C_P \leftrightarrow \frac{5\, m_p\, R}{Z \alpha} C_T
       \approx 3.8 \times 10^3\,\frac{A^{1/3}}{Z}C_T .
\label{CP_CT}
\end{equation}
The final coefficient connecting $C_P$ and $C_T$ in \eref{CP_CT}
differs by a factor of 6/5 from that obtained in~\cite{FlaKhr85}.

Given the results obtained in this work in the RPA approximation for
$d_{\rm at}^T$ and using~\eref{CP_CT} we can find the values of
$d_{\rm at}^P$.
These values are listed in~\tref{T:d3_1e} in the entry ``Rescaling (RPA)''.
As is seen from the table there is excellent agreement
between calculated and rescaled values. It is worth noting that
\eref{CP_CT} turns out to be insensitive to $Z$.
For the comparably light Xe and for heavy Ra the agreement of the
numerical results and the results obtained with use of~\eref{CP_CT} is
equally good. It means that \eref{CP_CT} works well for atoms
with different $Z$.
\begin{table}
\caption{The values of ${\bf d}_{\rm at}^P$
in units ($10^{-23} C_P\, \langle {\bm \sigma_N} \rangle |e|$ cm)
obtained in DHF and RPA approximations are presented.
The results are compared with other data.}
\label{T:d3_1e}
\begin{ruledtabular}
\begin{tabular}{lccccc}
                    & $^{129}$Xe & $^{171}$Yb & $^{199}$Hg & $^{211}$Rn & $^{225}$Ra \\
\hline
This work (DHF)     &    1.3     &   -2.4     &   -8.7     &    17.3    &    -13.0    \\
This work (RPA)     &    1.6     &  -11.5     &  -21.3     &    21.0    &    -63.7    \\
Rescaling (RPA)\footnotemark[1]
                    &    1.6     &  -11.3     &  -21.3     &    21.3    &    -64.7    \\
\end{tabular}
\end{ruledtabular}
\footnotemark[1]{These numbers are obtained using the correspondence
$C_P \leftrightarrow C_T$ given by~\eref{CP_CT} as explained in the text.}
\end{table}
\subsubsection{The Schiff moment}
The contribution to the atomic EDM due to the Schiff
moment $H^{SM}$ is naturally to determine as
${\bf d}_{\rm at}^{SM} = d_{\rm at}^{SM} ({\bf I}/{I})$,
where $d_{\rm at}^{SM}$ is given by
\begin{equation}
d_{\rm at}^{SM} = -2\,\frac{S}{B}\, \sum_{a,k} \frac
{\langle k||r|| a \rangle  \langle k|| r\, \rho(r)|| a \rangle}
{\varepsilon_k - \varepsilon_a} .
\label{Eq:SM}
\end{equation}
In explicit form the ME of the operator ($r\, \rho(r)$)
is given in Appendix~\ref{Ap:1}.

The results for the Schiff moment contribution to the atomic EDMs are presented in
Table~\ref{T:SM}. There is very good agreement with previous
calculations~\cite{DzuFlaGin02_EDM,DzuFlaGin07}. Few percent
difference for Hg and Ra is within the
accuracy of the calculations for these atoms.

A reason of the discrepancy with the result obtained in
Ref.~\cite{LatAngDas09} is unclear for us. The authors of this work
state that a possible reason of the difference between their result
and those obtained in Ref.~\cite{DzuFlaGin02_EDM} is due to electron
correlations which drastically change the final result.
However, according to our calculations, the correlations included in
the CI+MBPT approach change the results insignificantly, on the level of
10-15\% as compared to the RPA calculations.
The same conclusion was made in~Ref.\cite{DzuFlaGin02_EDM}.
Below we will return to this problem.
\begin{table}
\caption{The values of $d_{\rm at}^{SM}$
in units ($10^{-17}[S/(|e|\, {\rm fm}^3)]\, |e| {\rm cm}$)
obtained in DHF and RPA approximations are presented.
The results are compared with other data.}
\label{T:SM}
\begin{ruledtabular}
\begin{tabular}{lccccc}
                                 & $^{129}$Xe & $^{171}$Yb & $^{199}$Hg & $^{211}$Rn & $^{225}$Ra \\
\hline
This work (DHF)                  &    0.29    &  -0.42     &    -1.2    &  2.5       & -1.8   \\
This work (RPA)                  &    0.38    &  -1.9      &    -3.0    &  3.3       & -8.3   \\
Ref.\cite{DzuFlaGin02_EDM} (RPA) &    0.38    &            &    -2.8    &  3.3       & -8.5   \\
Ref.\cite{DzuFlaGin07} (RPA)     &            &  -1.9      &            &            &         \\
Ref.\cite{LatAngDas09} (RPA)     &            &            &    -5.07   &            &             \\
\end{tabular}
\end{ruledtabular}
\end{table}
\subsubsection{The electron EDM}

Usually in experiments searching for atomic EDMs atoms are
placed in an external electric field ${\bf E}_{\rm ext}$.
It leads to appearance of the interaction
${\bf d}_{\rm at} {\bf E}_{\rm ext} =
-d_e \gamma_5 {\bm \gamma}\,{\bf E}_{\rm ext}$.
The operator ($\gamma_5 {\bm \gamma}$) is P-even and T-odd.
If the hyperfine interaction, $H_{\rm hf}$, is accounted for, a contribution
to the atomic EDM caused by this operator appears already in the second order
of the perturbation theory and looks as follows
\begin{equation}
2 \sum_K \frac {\langle 0 |H_{\rm hf}| K \rangle
\langle K |d_e \gamma_5 {\bm \gamma}| 0 \rangle}
{E_0 - E_K} .
\end{equation}
But as it was shown in~\cite{FlaKhr85}, this contribution has
the magnitude $\sim d_e Z^3 \alpha^4 /m_p$ and is negligibly small.
We disregard it in this work.

For calculating the contribution of the electron EDM (described by the
operator $H_B$) to the atomic EDM we use the same approach as for
studying the $e$-$N$ interaction.
We denote the contribution to the atomic EDM due to $H_B$ as
${\bf d}_{\rm at}^B = d_{\rm at}^B ({\bf I}/{I})$.
Using \eref{Eq:d} and replacing $H \rightarrow H_B$,
after simple transformations we arrive at the following expression
\begin{equation}
d_{\rm at}^B = d_e\, \frac{\mu}{3 m_p c}\sum_{a,k} \frac
{\langle k||r|| a \rangle  \langle k||H_B^{\rm el}|| a \rangle}
{\varepsilon_k - \varepsilon_a} .
\label{Eq:dat_de}
\end{equation}
Here $H_B^{\rm el}$ is the electronic part of the operator $H_B$.
The explicit expression for the ME of the operator $H_B^{\rm el}$
is given in Appendix~\ref{Ap:1}.

The results obtained for this contribution to the atomic EDMs
is presented in \tref{T:HB}. Comparing our results
obtained for Xe and Hg with those of M{\aa}rtensson-Pendrill and \"{O}ster
we see very good agreement between them. It is seen that the RPA corrections
change the results obtained in the DHF approximation for Yb, Hg, and Ra
very significantly. For Yb and Ra the DHF and RPA results differ by a factor
of 5. This behavior is quit similar to what we found for $d_{\rm at}^T$,
$d_{\rm at}^P$, and $d_{\rm at}^{SM}$ (see Tables~\ref{T:d2_1e}, \ref{T:d3_1e},
and \ref{T:SM}). It is not surprisingly if we take into account the similar
nature of all these P,T-odd operators. To the best of our knowledge there are
no other available data for Yb, Rn, and Ra.

There is also the third order contribution to atomic EDM proportional $d_e$~\cite{FlaKhr85}
which is actually larger than the second order contribution discussed here. We will
consider it in a separate publication.
\begin{table}
\caption{The values of $d_{\rm at}^B$ in units ($d_e \times 10^{-4}$)
obtained in DHF and RPA approximations are presented.
The results are compared with other data.}
\label{T:HB}
\begin{ruledtabular}
\begin{tabular}{lccccc}
                             & $^{129}$Xe & $^{171}$Yb & $^{199}$Hg    & $^{211}$Rn & $^{225}$Ra \\
\hline
This work (DHF)              &    0.85   &   1.0     &   4.9         &   -11    &   -11    \\
Ref.\cite{MarOst87} (DHF)    &    0.86   &           &   5.1         &          &             \\
This work (RPA)              &    1.0    &   5.1     &  12.5         &   -13    &   -55     \\
Ref.\cite{MarOst87} (RPA)    &    1.05   &           &   13          &          &             \\
\end{tabular}
\end{ruledtabular}
\end{table}

\subsection{CI+MBPT}
\label{SubSec:nucl}
We use the CI+MBPT approximation for the calculations
for Yb, Hg, and Ra atoms, which have two $s$ electrons above closed shells.
For the noble gases such as Xe and Rn the RPA is known
to be good for describing their properties
(see, e.g.,~\cite{DzuFlaGin02}) and there is no need
to use the CI+MBPT for them.

We start the discussion of the properties of Yb, Hg, and Ra from the
results obtained for the low-lying energies of these atoms. In
\tref{Energy} we present the energy level values obtained in
the pure CI and the CI+MBPT approximation. As is seen from \tref{Energy}
the removal energies of the two $s$ electrons differ by $\sim$ 10\% from
the experimental values at the CI stage. An accounting for the MBPT
corrections leads to almost ideal (better than 0.1\%) agreement between
these quantities. The energies of the excited states (calculated relatively
to the ground state) are also noticeably improved at the CI+MBPT stage.
The differences between theoretical and experimental values do not exceed 3\%.
These results indicate the accuracy of wave functions produced at different
stages of the CI+MBPT method.
\begin{table}
\caption{The removal energies for both $6s$ electrons for Yb and Hg
and both $7s$ electrons for Ra are presented in the first row of each
respective atom. Energies of excited states are presented (in cm$^{-1}$)
in respect to the ground state. The results
are obtained in the CI and CI+MBPT approximations.}
\label{Energy}
\begin{ruledtabular}
\begin{tabular}{cccccc}
& Config. & Level & \multicolumn{1}{c}{CI}
&  \multicolumn{1}{c}{CI+MBPT}
&  \multicolumn{1}{c}{Experiment} \\
\hline
Yb & $6s^2$ & $^1S_0$   &  138795  & 148707  & 148712 \\
   & $6s6p$ & $^3P^o_0$ &   14368  &  17562  &  17288 \\
   & $6s6p$ & $^3P^o_1$ &   15029  &  18251  &  17992 \\
   & $6s6p$ & $^3P^o_2$ &   16537  &  19995  &  19710 \\
   & $6s6p$ & $^1P^o_1$ &   24215  &  25715  &  25068 \\[2mm]
Hg & $6s^2$ & $^1S_0$   &  213969  & 235409  & 235469 \\
   & $6s6p$ & $^3P^o_0$ &   30676  &  37537  &  37645 \\
   & $6s6p$ & $^3P^o_1$ &   32446  &  39375  &  39412 \\
   & $6s6p$ & $^3P^o_2$ &   36541  &  44405  &  44043 \\
   & $6s6p$ & $^1P^o_1$ &   48195  &  54116  &  54069 \\[2mm]
Ra & $7s^2$ & $^1S_0$   &  116301  & 124316  & 124416 \\
   & $7s7p$ & $^3P^o_0$ &   10706  &  13108  &  13078 \\
   & $7s7p$ & $^3P^o_1$ &   11587  &  14003  &  13999 \\
   & $7s7p$ & $^3P^o_2$ &   13892  &  16693  &  16686 \\
   & $7s7p$ & $^1P^o_1$ &   19011  &  20597  &  20716 \\
\end{tabular}
\end{ruledtabular}
\end{table}

The calculations of the atomic EDM ${\bf d}_{\rm at}$ in the CI+MBPT approah
is more complicated than in the RPA method.
Again we start from \eref{Eq:d} keeping in mind that the summation
in this equation is going over many-electron states.
Following \cite{DerJohSaf99,PorDer03} $d_{\rm at}$
can be divided into two parts,
\begin{equation}
d_{\rm at} = d^v_{\rm at} + d^{\rm core}_{\rm at}
\end{equation}
where $d^v_{\rm at}$ includes excitations of valence
electrons and $d^{\rm core}_{\rm at}$ includes
excitations of core electrons and a correction to
$d^{\rm core}_{\rm at}$ that appears
because of possible excitations of core electrons into
the closed valence $s$ shell, what is forbidden by the
Pauli principle. Note that the correction restoring
the Pauli principle is not small. In certain cases
it constitutes $\sim$ 50\% of the total core contribution.

With the wave functions obtained from Eq.~(\ref{Phi}), the
valence part $d^v_{\rm at}$ is computed with the
Sternheimer~\cite{Ste50} or Dalgarno-Lewis \cite{DalLew55} method
implemented in the CI+MBPT+RPA framework. Given the wave
function $|0\rangle$ and its energy $E_0$, we find an intermediate-state wave
function $\Phi_{\rm in}$ from the inhomogeneous equation
\begin{eqnarray}
|\Phi_{\rm in} \rangle & = & \frac{1}{H_{\mathrm{eff}} -E_0}\,
\sum_K | K \rangle \langle K | (r_z)_{\mathrm{eff}} |0\rangle \nonumber \\
&=&  \frac{1}{H_{\mathrm{eff}}-E_0}\,
(r_z)_{\mathrm{eff}} |0 \rangle .
\label{Phi_in}
\end{eqnarray}
Given $\Phi_{\rm in}$ we can be compute $d^v_{\rm at}$ as
\begin{equation}
d^v_{\rm at} = 2\, \langle 0 |(r_z)_{\mathrm{eff}}| \Phi_{\rm in} \rangle.
\label{dv}
\end{equation}
An additional contribution $d^{\rm core}_{\rm at}$ coming from particle-hole
excitations of the core is incorporated in the frame of the RPA approach
discussed above.

In \tref{T:allEDM} we list the results obtained for the valence
and core contributions to the EDM of the atoms. For completeness and comparison
we also present the values obtained in the RPA calculations.
\begin{table}
\caption{The valence and core and contributions to ${\bf d}_{\rm at}^T$
(in $10^{-20} C_T\, \langle {\bm \sigma_N} \rangle |e|$ cm),
${\bf d}_{\rm at}^P$ (in $10^{-23} C_P\, \langle {\bm \sigma_N} \rangle |e|$ cm),
$d^{\rm SM}_{\rm at}$ (in $10^{-17}  [S/(|e|\, {\rm fm}^3)]\, |e|\, {\rm cm}$), and
$d^B_{\rm at}$ (in $10^{-4} d_e$) are presented.
The entry ``Total'' means the sum of the valence and core contributions.
The results obtained in the RPA method ($V^N$ approximation) are given for
comparison. We denote this entry as ``RPA''.}
\label{T:allEDM}
\begin{ruledtabular}
\begin{tabular}{lccccc}
&& \multicolumn{1}{c}{${\bf d}_{\rm at}^T$}
 & \multicolumn{1}{c}{${\bf d}_{\rm at}^P$}
 & \multicolumn{1}{c}{$d^{\rm SM}_{\rm at}$}
 & \multicolumn{1}{c}{$d^B_{\rm at}$}\\
\hline
$^{171}$Yb &  val.    &   -4.24    &   -14.2   &   -2.43    &   6.24   \\
           &  core    &    0.54    &     1.8   &    0.31    &  -0.79   \\
           & Total    &   -3.70    &   -12.4   &   -2.12    &   5.45   \\
           & RPA      &   -3.37    &   -10.9   &   -1.95    &   5.05   \\[2mm]
$^{199}$Hg &  val.    &   -5.71    &   -20.5   &   -2.95    &   12.0   \\
           &  core    &    0.59    &     2.1   &    0.32    &   -1.3   \\
           & Total    &   -5.12    &   -18.4   &   -2.63    &   10.7   \\
           & RPA      &   -5.89    &   -20.7   &   -2.99    &   12.3   \\[2mm]
$^{225}$Ra &  val.    &   -20.6    &   -75.0   &   -10.25   &  -65.2   \\
           &  core    &     3.0    &    10.8   &     1.41   &    9.5   \\
           & Total    &   -17.6    &   -64.2   &    -8.84   &  -55.7   \\
           & RPA      &   -16.7    &   -61.0   &    -8.27   &  -53.3   \\
\end{tabular}
\end{ruledtabular}
\end{table}
As is seen from the table the largest differences between the results
obtained in the RPA and CI+MBPT methods occur for Hg. But even
in this case these differences do not exceed 15\%. Taking into account the similar nature
of the P,T-odd operators considered in this work we can expect that the
relative difference between the results found in both approximations
would be approximately the same for all operators for a given atom. As we see
from \tref{T:allEDM} this condition is fulfilled. This is a good
consistency test of the calculations.

In this way we arrive at the conclusion that the electron correlations do not
affect the final results too much. Based on this observation we
estimate the accuracy of the values ``Total'' listed in \tref{T:allEDM}
is at the level of 15-20\%. Similar accuracy is expected for Xe and Rn atoms.
\section{Shell model of the nucleus}
\label{Sec:nucl}

In certain cases it is possible to impose additionally
the constraints on the strength of the couplings for protons and neutrons.
Below we will obtain such constraints for a simple case of the
spherical shell model of a nucleus. This model describes well the
nuclei of $^{129}$Xe and $^{199}$Hg. The former contains a neutron
in the $s_{1/2}$ state and the latter contains a neutron in the $p_{1/2}$
state above the closed shells. Although the nuclei in the atoms
have a valence neutron it is possible to deduce also the proton contribution to the total nuclear spin
using the information on the nuclear magnetic moments. If
we assume that the magnetic moment of the nucleus is composed entirely from
the spin magnetic moment of the valence neutron and spin magnetism of
polarized nuclear core, then
\begin{eqnarray}
\mu = \mu_n \langle \sigma_{z}^{(n)} \rangle + \mu_p \langle \sigma_{z}^{(p)} \rangle \\
\langle \sigma_{z}^{(n)} \rangle + \langle \sigma_{z}^{(p)}
\rangle = \langle \sigma_{z}^{(0)} \rangle. \nonumber
\label{simplespin}
\end{eqnarray}
Here $\mu_p \approx 2.793$ and $\mu_n \approx -1.913$ are the magentic moments of the
proton and neutron expressed in nuclear magnetons.
Numerical estimates show that the main contribution to the nuclear magnetic moments $\mu$
of $^{129}$Xe and $^{199}$Hg comes from the neutron and proton spin contributions.
This is due to that the neutron
orbital contribution is zero and the proton orbital contribution
(for the orbitals with a low orbital momentum) is small in comparison
with its spin contribution.
Neglecting the spin-orbit interaction leads to conservation of the total spin
which is equal to the average spin of the neutron above the unpolarized core
$\langle \sigma_{z}^{(0)} \rangle$.
Taking into account that in the spherical shell model the nuclear spin ${\bf I}$
is determined by the total momentum of the unpaired nucleon we can write
\begin{equation}
\langle {\bm \sigma}_N \rangle = \langle \sigma_{z}^{(0)} \rangle \, {\bf I}/I ,
\end{equation}
with
$$
\langle \sigma_{z}^{(0)} \rangle =
\left\lbrace
\begin{array}{l}
\,\,\quad 1,\qquad \quad I = l_I + 1/2  \\
-I/(I+1),\, I = l_I - 1/2
\end{array}
\right. ,
$$
where $l_I$ is the orbital quantum number of the valence nucleon.
Using Eq.~(\ref{simplespin}), we determine  $\langle \sigma_{z}^{(n)} \rangle$
and $\langle \sigma_{z}^{(p)} \rangle$ for observationally relevant cases of
$^{129}$Xe and $^{129}$Hg as shown in \tref{nucl}.
\begin{table}
\caption{Composition of the nuclear spin.}
\label{nucl}
\begin{ruledtabular}
\begin{tabular}{ccccc}
  Nucleus   & Neutron state & $\langle \sigma_{z}^{(0)} \rangle$ &
 $\langle \sigma_{z}^{(n)} \rangle$ &  $\langle \sigma_{z}^{(p)} \rangle$ \\
\hline
 $^{129}$Xe &   $s_{1/2}$  &  1     &  0.76 &  0.24 \\
 $^{199}$Hg &   $p_{1/2}$  & -1/3   & -0.31 & -0.03 \\
\end{tabular}
\end{ruledtabular}
\end{table}

As is seen from \tref{nucl} the contribution of the proton spin into the total
nuclear spin of $^{129}$Xe is as high as 30\%, and therefore the proton couplings
$C^p_T$ and $C^p_P$ (see the equations below Eq.~(\ref{Eq:rho}))
are also limited in the experiments searching for the EDM.
For $^{199}$Hg the limit on the proton couplings is 10 times weaker than the limit on the
neutron couplings.

Note that the simple shell model of a nucleus considered above is hardly applicable
to the nuclei of $^{171}$Yb, $^{211}$Rn, and $^{225}$Ra. The problem is that the nucleus
of $^{171}$Yb is quadrupole-deformed and the nuclei of $^{211}$Rn and $^{225}$Ra
are octupole-deformed. For this reason more sophisticated nuclear models are required
for a proper description of these nuclei.
\section{Conclusion}
\label{Sec:concl}
In conclusion, we have carried out calculations of different
contributions to the EDMs of $^{129}$Xe, $^{171}$Yb, $^{199}$Hg,
$^{211}$Rn, and $^{225}$Ra by the RPA and the CI+MBPT methods.
Two of these contributions are due to
tensor-pseudotensor and scalar-pseudoscalar $e$-$N$
P,T-odd interactions, and two more contributions are caused by
the nuclear Schiff moment and the electron EDM.
The recommended values for noble gases Xn and Ra are based on
the results obtained in the RPA calculations while for
Yb, Hg, and Ra they based on the calculations
carried out in the frame of CI+MBPT+RPA approach.
These numbers are gathered in \tref{T:final}.
\begin{table*}
\caption{The recommended values of the contributions to atomic EDM.}
\label{T:final}
\begin{ruledtabular}
\begin{tabular}{lccccc}
                             & $^{129}$Xe & $^{171}$Yb & $^{199}$Hg    & $^{211}$Rn & $^{225}$Ra \\
\hline
${\bf d}_{\rm at}^T (10^{-20} C_T\, \langle {\bm \sigma_N} \rangle |e|$ cm)
                             &    0.57    &   -3.7     &   -5.1   &   5.6     &   -18        \\
${\bf d}_{\rm at}^P (10^{-23} C_P\, \langle {\bm \sigma_N} \rangle |e|$ cm)
                             &    1.6     &   -12      &   -18    &   21      &   -64       \\
$d^{\rm SM}_{\rm at} (10^{-17}  [S/(|e|\, {\rm fm}^3)] |e|\, {\rm cm})$
                             &    0.38    &   -2.1     &   -2.6   &   3.3     &   -8.8       \\
$d^B_{\rm at} (10^{-4} d_e$)
                             &    1.0     &    5.5     &   11     &   -13     &   -56        \\
\end{tabular}
\end{ruledtabular}
\end{table*}

Finally, using the results obtained in this work for atomic mercury and recently
obtained new upper bound
$|d_{\rm at}(^{199}\!{\rm Hg})| < 3.1 \times 10^{-29}\, |e|\,{\rm cm}$~\cite{GriSwaLof09}
we are able to find limits on $CP$ violating parameters $C_T$, $C_P$, $S$.
We do not put a constraint on the coupling constant $d_e$ because
the contribution of the electric dipole electron moment to the atomic EDM in the third order
of the perturbation theory is known to be $\sim$ 10 times larger than its contribution in the
second order~\cite{FlaKhr85}. It means that a dominating contribution from the electron EDM
still has to be considered. This is a subject of another work.
Applying again the spherical shell model to $^{199}$Hg
and, respectively, having $\langle {\bm \sigma_N} \rangle = -1/3\,({\bf I}/I)$
we arrive at the numbers listed in \tref{T:CP}.
\begin{table}
\caption{Limits on $CP$ violating parameters $C_T$, $C_P$,
and $S$ based on new experimental limit for
$|d_{\rm at}(^{199}\!{\rm Hg})| < 3.1 \times 10^{-29}\, |e|$ cm~\cite{GriSwaLof09}}
\label{T:CP}
\begin{ruledtabular}
\begin{tabular}{lc}
  Parameter             &  Limit               \\
\hline
   $C_T$                &   1.9  $\times 10^{-9}$ \\
   $C_P$                &   5.2  $\times 10^{-7}$ \\
$S\, (|e|\,{\rm fm}^3)$ &   1.2  $\times 10^{-12}$ \\
\end{tabular}
\end{ruledtabular}
\end{table}



We would like to thank M. Kozlov for useful
remarks. This work was supported by Australian Research Council.
S.G.P. was also supported in part by the
Russian Foundation for Basic Research under Grants No.~07-02-00210-a
and No.~08-02-00460-a.

\appendix
\section{}
\label{Ap:1}
To calculate the MEs of the P,T-operators we define the one-electron
wave function $|a\rangle \equiv \psi_a({\bf r})$ as follows
\begin{equation}
\psi_a ({\bf r}) =
\left(
\begin{array}{l}
f_a(r)\,\, \Omega_{j_a l_a m_a} ({\bf n}) \\
i g_a(r)\, \Omega_{j_a \tilde{l}_a m_a}({\bf n})
\end{array}
\right) ,
\end{equation}
where $\tilde{l}_a = 2j_a-l_a$.

The MEs of the P,T-odd operators characterizing
the tensor-pseudotensor and pseudoscalar-scalar interaction
are given by the following expressions
\begin{eqnarray}
&&\langle n_b \kappa_b || i \gamma\, \rho(r)|| n_a \kappa_a \rangle =
\langle \kappa_b ||C_1|| \kappa_a \rangle \times  \\
&& \int_0^\infty \!
\left\{f_b g_a (\kappa_a\!-\kappa_b\!+1) + \!
f_a g_b (\kappa_b\!-\kappa_a\!+1)\! \right\} \rho(r) r^2 dr , \nonumber
\end{eqnarray}
\begin{eqnarray}
&&\langle n_b \kappa_b ||n\, (d\rho/dr) \gamma_0|| n_a \kappa_a \rangle =
 \langle \kappa_b ||C_1|| \kappa_a \rangle
\nonumber \\ && \times \int_0^\infty \!
 \left( g_b g_a - f_b f_a\right) r^2 \, \frac{d\rho}{dr} dr .
\label{Ap:rho}
\end{eqnarray}
If the nuclear density distribution is given by \eref{Eq:rho}
(see the main text) \eref{Ap:rho}  is further simplified
leading to
\begin{eqnarray}
&&\langle n_b \kappa_b ||n\, (d\rho/dr) \gamma_0|| n_a \kappa_a \rangle = \nonumber \\
&& \langle \kappa_b ||C_1|| \kappa_a \rangle
 \frac{3}{4\pi} \left( \frac{g_b g_a - f_b f_a}{r} \right)_{r=R} ,
\end{eqnarray}
where spherical harmonics $C_{lm}$ are defined as
\begin{equation}
 C_{lm}({\bf n}) = \sqrt{\frac{4\pi}{2l+1}}\, Y_{lm}({\bf n}) ,
\end{equation}
$\kappa = (l-j)(2j+1)$
and the reduced ME $\langle \kappa_b ||C_1|| \kappa_a \rangle$ is
given by
\begin{eqnarray}
&&\langle \kappa_b ||C_1|| \kappa_a \rangle = \xi(l_b+l_a+1) \times  \\
&& (-1)^{j_b+1/2} \sqrt{(2j_a+1)(2j_b+1)}
\left( \!
\begin{array}{ccc}
 j_b & j_a & 1 \\
-1/2 & 1/2 & 0
\end{array}
\! \right) ,  \nonumber
\end{eqnarray}
where
$ \xi(x) =
\left\lbrace
\begin{array}{l}
1,\, {\rm if}\,\, x\,\, {\rm is\,\, even}  \\
0,\, {\rm if}\,\, x\,\, {\rm is\,\, odd}
\end{array}
\right. $ .

The reduced ME characterizing the nuclear Schiff moment is given by
\begin{eqnarray}
&&\langle n_b \kappa_b ||r \, \rho(r) || n_a \kappa_a \rangle =
\langle \kappa_b ||C_1|| \kappa_a \rangle \times  \\
&& \int_0^\infty \!
\left\{g_b g_a  + \! f_b f_a \! \right\} r^3 \, \rho(r) dr . \nonumber
\end{eqnarray}

The reduced ME of the operator $H_B^{\rm el}$
can be represented as
\begin{eqnarray}
&&\langle n_b \kappa_b ||H_B^{\rm el}|| n_a \kappa_a \rangle =
-\langle \kappa_b ||C_1|| \kappa_a \rangle \times  \\
&& \left[ \! \int_R^\infty \!
\left\{ f_b g_a (\kappa_b\!-\kappa_a\!+2)\! + \!
f_a g_b (\kappa_a\!-\kappa_b\!+2) \right\} \frac{dr}{r} - \right. \nonumber \\
&& 2\! \left. \! \int_0^R \!
\left\{ f_b g_a (\kappa_b\!-\kappa_a\!-1)\! + \!
f_a g_b (\kappa_a\!-\kappa_b\!-1) \right\} \frac{r^2 dr}{R^3} \right] .\nonumber
\end{eqnarray}


\begin{thebibliography}{29}
\expandafter\ifx\csname natexlab\endcsname\relax\def\natexlab#1{#1}\fi
\expandafter\ifx\csname bibnamefont\endcsname\relax
  \def\bibnamefont#1{#1}\fi
\expandafter\ifx\csname bibfnamefont\endcsname\relax
  \def\bibfnamefont#1{#1}\fi
\expandafter\ifx\csname citenamefont\endcsname\relax
  \def\citenamefont#1{#1}\fi
\expandafter\ifx\csname url\endcsname\relax
  \def\url#1{\texttt{#1}}\fi
\expandafter\ifx\csname urlprefix\endcsname\relax\def\urlprefix{URL }\fi
\providecommand{\bibinfo}[2]{#2}
\providecommand{\eprint}[2][]{\url{#2}}

\bibitem[{\citenamefont{Khriplovich and Lamoreaux}(1997)}]{KhrLam97}
\bibinfo{author}{\bibfnamefont{I.~B.} \bibnamefont{Khriplovich}}
  \bibnamefont{and} \bibinfo{author}{\bibfnamefont{S.~K.}
  \bibnamefont{Lamoreaux}}, \emph{\bibinfo{title}{CP violation without
  strangeness. Electric dipole moments of particles, atoms, and molecules.}}
  (\bibinfo{publisher}{Springer}, \bibinfo{address}{Berlin},
  \bibinfo{year}{1997}).

\bibitem[{\citenamefont{Huber et~al.}(2007)\citenamefont{Huber, Pospelov, and
  Ritz}}]{HubPosRit07}
\bibinfo{author}{\bibfnamefont{S.~J.} \bibnamefont{Huber}},
  \bibinfo{author}{\bibfnamefont{M.}~\bibnamefont{Pospelov}}, \bibnamefont{and}
  \bibinfo{author}{\bibfnamefont{A.}~\bibnamefont{Ritz}},
  \bibinfo{journal}{Phys. Rev. D} \textbf{\bibinfo{volume}{75}},
  \bibinfo{pages}{036006} (\bibinfo{year}{2007}).

\bibitem[{\citenamefont{Pospelov and Ritz}(2005)}]{PosRit05}
\bibinfo{author}{\bibfnamefont{M.}~\bibnamefont{Pospelov}} \bibnamefont{and}
  \bibinfo{author}{\bibfnamefont{A.}~\bibnamefont{Ritz}},
  \bibinfo{journal}{Ann. Phys. (NY)} \textbf{\bibinfo{volume}{318}},
  \bibinfo{pages}{119} (\bibinfo{year}{2005}).

\bibitem[{\citenamefont{Griffith et~al.}(2009)\citenamefont{Griffith, Swallows,
  Loftus, Romalis, Heckel, and Fortson}}]{GriSwaLof09}
\bibinfo{author}{\bibfnamefont{W.~C.} \bibnamefont{Griffith}},
  \bibinfo{author}{\bibfnamefont{M.~D.} \bibnamefont{Swallows}},
  \bibinfo{author}{\bibfnamefont{T.~H.} \bibnamefont{Loftus}},
  \bibinfo{author}{\bibfnamefont{M.~V.} \bibnamefont{Romalis}},
  \bibinfo{author}{\bibfnamefont{B.~R.} \bibnamefont{Heckel}},
  \bibnamefont{and} \bibinfo{author}{\bibfnamefont{E.~N.}
  \bibnamefont{Fortson}}, \bibinfo{journal}{Phys. Rev. Lett.}
  \textbf{\bibinfo{volume}{102}}, \bibinfo{pages}{101601}
  (\bibinfo{year}{2009}).

\bibitem[{\citenamefont{Ginges and Flambaum}(2004)}]{GinFla04}
\bibinfo{author}{\bibfnamefont{J.~S.~M.} \bibnamefont{Ginges}}
  \bibnamefont{and} \bibinfo{author}{\bibfnamefont{V.~V.}
  \bibnamefont{Flambaum}}, \bibinfo{journal}{Phys. Rep.}
  \textbf{\bibinfo{volume}{397}}, \bibinfo{pages}{63} (\bibinfo{year}{2004}).

\bibitem[{\citenamefont{Flambaum and Ginges}(2002)}]{FlaGin02}
\bibinfo{author}{\bibfnamefont{V.~V.}~\bibnamefont{Flambaum}} \bibnamefont{and}
  \bibinfo{author}{\bibfnamefont{J.~S.~M.}~\bibnamefont{Ginges}},
  \bibinfo{journal}{Phys. Rev. A} \textbf{\bibinfo{volume}{65}},
  \bibinfo{pages}{032113} (\bibinfo{year}{2002}).

\bibitem[{\citenamefont{Dzuba et~al.}(2002{\natexlab{a}})\citenamefont{Dzuba,
  Flambaum, Ginges, and Kozlov}}]{DzuFlaGin02_EDM}
\bibinfo{author}{\bibfnamefont{V.~A.} \bibnamefont{Dzuba}},
  \bibinfo{author}{\bibfnamefont{V.~V.} \bibnamefont{Flambaum}},
  \bibinfo{author}{\bibfnamefont{{\rm J. S. M}.}~\bibnamefont{Ginges}},
  \bibnamefont{and} \bibinfo{author}{\bibfnamefont{M.~G.}
  \bibnamefont{Kozlov}}, \bibinfo{journal}{Phys. Rev. A}
  \textbf{\bibinfo{volume}{66}}, \bibinfo{pages}{012111}
  (\bibinfo{year}{2002}{\natexlab{a}}).

\bibitem[{\citenamefont{Dzuba et~al.}(2007)\citenamefont{Dzuba, Flambaum, and
  Ginges}}]{DzuFlaGin07}
\bibinfo{author}{\bibfnamefont{V.~A.} \bibnamefont{Dzuba}},
  \bibinfo{author}{\bibfnamefont{V.~V.} \bibnamefont{Flambaum}},
  \bibnamefont{and} \bibinfo{author}{\bibfnamefont{{\rm J. S.
  M}.}~\bibnamefont{Ginges}}, \bibinfo{journal}{Phys. Rev. A}
  \textbf{\bibinfo{volume}{76}}, \bibinfo{pages}{034501}
  (\bibinfo{year}{2007}).

\bibitem[{\citenamefont{Kozlov et~al.}(1996)\citenamefont{Kozlov, Porsev, and
  Flambaum}}]{KozPorFla96}
\bibinfo{author}{\bibfnamefont{M.~G.} \bibnamefont{Kozlov}},
  \bibinfo{author}{\bibfnamefont{S.~G.} \bibnamefont{Porsev}},
  \bibnamefont{and} \bibinfo{author}{\bibfnamefont{V.~V.}
  \bibnamefont{Flambaum}}, \bibinfo{journal}{J. \ Phys. \ B}
  \textbf{\bibinfo{volume}{29}}, \bibinfo{pages}{689} (\bibinfo{year}{1996}).

\bibitem[{\citenamefont{Bogdanovich}(1991)}]{Bog91}
\bibinfo{author}{\bibfnamefont{P.}~\bibnamefont{Bogdanovich}},
  \bibinfo{journal}{Lith. Phys. J.} \textbf{\bibinfo{volume}{31}},
  \bibinfo{pages}{79} (\bibinfo{year}{1991}).

\bibitem[{\citenamefont{Johnson et~al.}(1988)\citenamefont{Johnson, Blundell,
  and Sapirstein}}]{JohBluSap88}
\bibinfo{author}{\bibfnamefont{W.~R.} \bibnamefont{Johnson}},
  \bibinfo{author}{\bibfnamefont{S.~A.} \bibnamefont{Blundell}},
  \bibnamefont{and}
  \bibinfo{author}{\bibfnamefont{J.}~\bibnamefont{Sapirstein}},
  \bibinfo{journal}{Phys.\ Rev.\ A} \textbf{\bibinfo{volume}{37}},
  \bibinfo{pages}{307} (\bibinfo{year}{1988}).

\bibitem[{\citenamefont{Dzuba et~al.}(1996)\citenamefont{Dzuba, Flambaum, and
  Kozlov}}]{DzuFlaKoz96b}
\bibinfo{author}{\bibfnamefont{V.~A.} \bibnamefont{Dzuba}},
  \bibinfo{author}{\bibfnamefont{V.~V.} \bibnamefont{Flambaum}},
  \bibnamefont{and} \bibinfo{author}{\bibfnamefont{M.~G.}
  \bibnamefont{Kozlov}}, \bibinfo{journal}{Phys.\ Rev.\ A}
  \textbf{\bibinfo{volume}{54}}, \bibinfo{pages}{3948} (\bibinfo{year}{1996}).

\bibitem[{\citenamefont{Dzuba et~al.}(1998)\citenamefont{Dzuba, Kozlov, Porsev,
  and Flambaum}}]{DzuKozPor98}
\bibinfo{author}{\bibfnamefont{V.~A.} \bibnamefont{Dzuba}},
  \bibinfo{author}{\bibfnamefont{M.~G.} \bibnamefont{Kozlov}},
  \bibinfo{author}{\bibfnamefont{S.~G.} \bibnamefont{Porsev}},
  \bibnamefont{and} \bibinfo{author}{\bibfnamefont{V.~V.}
  \bibnamefont{Flambaum}}, \bibinfo{journal}{Zh. \ Eksp. \ Teor. \ Fiz.}
  \textbf{\bibinfo{volume}{114}}, \bibinfo{pages}{1636} (\bibinfo{year}{1998}),
  \bibinfo{note}{[Sov. \ Phys.--JETP {\bf 87} 885, (1998)]}.

\bibitem[{\citenamefont{Porsev et~al.}(1999{\natexlab{a}})\citenamefont{Porsev,
  Rakhlina, and Kozlov}}]{PorRakKoz99P}
\bibinfo{author}{\bibfnamefont{S.~G.} \bibnamefont{Porsev}},
  \bibinfo{author}{\bibfnamefont{{\rm Yu}.~G.} \bibnamefont{Rakhlina}},
  \bibnamefont{and} \bibinfo{author}{\bibfnamefont{M.~G.}
  \bibnamefont{Kozlov}}, \bibinfo{journal}{Phys. Rev. A}
  \textbf{\bibinfo{volume}{60}}, \bibinfo{pages}{2781}
  (\bibinfo{year}{1999}{\natexlab{a}}).

\bibitem[{\citenamefont{Porsev et~al.}(1999{\natexlab{b}})\citenamefont{Porsev,
  Rakhlina, and Kozlov}}]{PorRakKoz99J}
\bibinfo{author}{\bibfnamefont{S.~G.} \bibnamefont{Porsev}},
  \bibinfo{author}{\bibfnamefont{{\rm Yu}.~G.} \bibnamefont{Rakhlina}},
  \bibnamefont{and} \bibinfo{author}{\bibfnamefont{M.~G.}
  \bibnamefont{Kozlov}}, \bibinfo{journal}{J. Phys. B}
  \textbf{\bibinfo{volume}{32}}, \bibinfo{pages}{1113}
  (\bibinfo{year}{1999}{\natexlab{b}}).

\bibitem[{\citenamefont{Kolb et~al.}(1982)\citenamefont{Kolb, Johnson, and
  Shorer}}]{KolJohSho82}
\bibinfo{author}{\bibfnamefont{D.}~\bibnamefont{Kolb}},
  \bibinfo{author}{\bibfnamefont{W.~R.} \bibnamefont{Johnson}},
  \bibnamefont{and} \bibinfo{author}{\bibfnamefont{P.}~\bibnamefont{Shorer}},
  \bibinfo{journal}{Phys. Rev. A} \textbf{\bibinfo{volume}{26}},
  \bibinfo{pages}{19} (\bibinfo{year}{1982}).

\bibitem[{\citenamefont{Johnson et~al.}(1983)\citenamefont{Johnson, Kolb, and
  Huang}}]{JohKolHua83}
\bibinfo{author}{\bibfnamefont{W.~R.} \bibnamefont{Johnson}},
  \bibinfo{author}{\bibfnamefont{D.}~\bibnamefont{Kolb}}, \bibnamefont{and}
  \bibinfo{author}{\bibfnamefont{{\rm K.-N}.}~\bibnamefont{Huang}},
  \bibinfo{journal}{At. Data Nucl. Data Tables} \textbf{\bibinfo{volume}{28}},
  \bibinfo{pages}{333} (\bibinfo{year}{1983}).

\bibitem[{\citenamefont{M{\aa}rtensson-Pendrill}(1985)}]{Mar85}
\bibinfo{author}{\bibfnamefont{A.~M.}~\bibnamefont{M{\aa}rtensson-Pendrill}},
  \bibinfo{journal}{Phys. Rev. Lett.} \textbf{\bibinfo{volume}{54}},
  \bibinfo{pages}{1153} (\bibinfo{year}{1985}).

\bibitem[{\citenamefont{Latha et~al.}(2008)\citenamefont{Latha, Angom,
  Chaudhuri, Das, and Mukherjee}}]{LatAngCha08}
\bibinfo{author}{\bibfnamefont{K.~{\rm V. P}.} \bibnamefont{Latha}},
  \bibinfo{author}{\bibfnamefont{D.}~\bibnamefont{Angom}},
  \bibinfo{author}{\bibfnamefont{R.~J.} \bibnamefont{Chaudhuri}},
  \bibinfo{author}{\bibfnamefont{B.~P.} \bibnamefont{Das}}, \bibnamefont{and}
  \bibinfo{author}{\bibfnamefont{D.}~\bibnamefont{Mukherjee}},
  \bibinfo{journal}{J. Phys. B} \textbf{\bibinfo{volume}{41}},
  \bibinfo{pages}{035005} (\bibinfo{year}{2008}).

\bibitem[{\citenamefont{Sushkov et~al.}(1984)\citenamefont{Sushkov, Flambaum,
  and Khriplovich}}]{SusFlaKhr84}
\bibinfo{author}{\bibfnamefont{O.~P.} \bibnamefont{Sushkov}},
  \bibinfo{author}{\bibfnamefont{V.~V.} \bibnamefont{Flambaum}},
  \bibnamefont{and} \bibinfo{author}{\bibfnamefont{I.~B.}
  \bibnamefont{Khriplovich}}, \bibinfo{journal}{Zh. Eksp. Theor. Fiz.}
  \textbf{\bibinfo{volume}{87}}, \bibinfo{pages}{1521} (\bibinfo{year}{1984}),
  \bibinfo{note}{[Sov. Phys. JETP {\bf 60}, 873 (1984)]}.

\bibitem[{\citenamefont{Dzuba et~al.}(1985)\citenamefont{Dzuba, Flambaum, and
  Silvestrov}}]{DzuFlaSil85}
\bibinfo{author}{\bibfnamefont{V.~A.} \bibnamefont{Dzuba}},
  \bibinfo{author}{\bibfnamefont{V.~V.} \bibnamefont{Flambaum}},
  \bibnamefont{and} \bibinfo{author}{\bibfnamefont{P.~G.}
  \bibnamefont{Silvestrov}}, \bibinfo{journal}{Phys. Lett. B}
  \textbf{\bibinfo{volume}{154B}}, \bibinfo{pages}{93} (\bibinfo{year}{1985}).

\bibitem[{\citenamefont{Flambaum and Khriplovich}(1985)}]{FlaKhr85}
\bibinfo{author}{\bibfnamefont{V.~V.} \bibnamefont{Flambaum}} \bibnamefont{and}
  \bibinfo{author}{\bibfnamefont{I.~B.} \bibnamefont{Khriplovich}},
  \bibinfo{journal}{Zh. Eksp. Theor. Fiz.} \textbf{\bibinfo{volume}{89}},
  \bibinfo{pages}{1505} (\bibinfo{year}{1985}).

\bibitem[{\citenamefont{Latha et~al.}(2009)\citenamefont{Latha, Angom, Das, and
  Mukherjee}}]{LatAngDas09}
\bibinfo{author}{\bibfnamefont{K.~{\rm V. P}.} \bibnamefont{Latha}},
  \bibinfo{author}{\bibfnamefont{D.}~\bibnamefont{Angom}},
  \bibinfo{author}{\bibfnamefont{B.~P.} \bibnamefont{Das}}, \bibnamefont{and}
  \bibinfo{author}{\bibfnamefont{D.}~\bibnamefont{Mukherjee}}
  (\bibinfo{year}{2009}), \bibinfo{note}{arXiv:0902.4790v1}.

\bibitem[{\citenamefont{M{\aa}rtensson-Pendrill and
  \"{O}ster}(1987)}]{MarOst87}
\bibinfo{author}{\bibfnamefont{A.~M.}~\bibnamefont{M{\aa}rtensson-Pendrill}}
  \bibnamefont{and}
  \bibinfo{author}{\bibfnamefont{P.}~\bibnamefont{\"{O}ster}},
  \bibinfo{journal}{Phys. Scr.} \textbf{\bibinfo{volume}{36}},
  \bibinfo{pages}{444} (\bibinfo{year}{1987}).

\bibitem[{\citenamefont{Dzuba et~al.}(2002{\natexlab{b}})\citenamefont{Dzuba,
  Flambaum, and Ginges}}]{DzuFlaGin02}
\bibinfo{author}{\bibfnamefont{V.~A.}~\bibnamefont{Dzuba}},
  \bibinfo{author}{\bibfnamefont{V.~V.}~\bibnamefont{Flambaum}}, \bibnamefont{and}
  \bibinfo{author}{\bibfnamefont{J.~S.~M.}~\bibnamefont{Ginges}},
  \bibinfo{journal}{Phys. Rev. D} \textbf{\bibinfo{volume}{66}},
  \bibinfo{pages}{076013} (\bibinfo{year}{2002}{\natexlab{b}}).

\bibitem[{\citenamefont{Derevianko et~al.}(1999)\citenamefont{Derevianko,
  Johnson, Safronova, and Babb}}]{DerJohSaf99}
\bibinfo{author}{\bibfnamefont{A.}~\bibnamefont{Derevianko}},
  \bibinfo{author}{\bibfnamefont{W.~R.} \bibnamefont{Johnson}},
  \bibinfo{author}{\bibfnamefont{M.~S.} \bibnamefont{Safronova}},
  \bibnamefont{and} \bibinfo{author}{\bibfnamefont{J.~F.} \bibnamefont{Babb}},
  \bibinfo{journal}{Phys.\ Rev.\ Lett.} \textbf{\bibinfo{volume}{82}},
  \bibinfo{pages}{3589} (\bibinfo{year}{1999}).

\bibitem[{\citenamefont{Porsev and Derevianko}(2003)}]{PorDer03}
\bibinfo{author}{\bibfnamefont{S.~G.} \bibnamefont{Porsev}} \bibnamefont{and}
  \bibinfo{author}{\bibfnamefont{A.}~\bibnamefont{Derevianko}},
  \bibinfo{journal}{J. Chem. Phys.} \textbf{\bibinfo{volume}{119}},
  \bibinfo{pages}{844} (\bibinfo{year}{2003}).

\bibitem[{\citenamefont{Sternheimer}(1950)}]{Ste50}
\bibinfo{author}{\bibfnamefont{R.~M.} \bibnamefont{Sternheimer}},
  \bibinfo{journal}{Phys. Rev.} \textbf{\bibinfo{volume}{80}},
  \bibinfo{pages}{102} (\bibinfo{year}{1950}).

\bibitem[{\citenamefont{Dalgarno and Lewis}(1955)}]{DalLew55}
\bibinfo{author}{\bibfnamefont{A.}~\bibnamefont{Dalgarno}} \bibnamefont{and}
  \bibinfo{author}{\bibfnamefont{J.~T.} \bibnamefont{Lewis}},
  \bibinfo{journal}{Proc.\ Roy.\ Soc.} \textbf{\bibinfo{volume}{233}},
  \bibinfo{pages}{70} (\bibinfo{year}{1955}).

\end{thebibliography}
\end{document}